# EdgeMORE: improving resource allocation with multiple options from tenants

Andrea Araldo*, Alessandro Di Stefano† and Antonella Di Stefano†

* Télécom SudParis
Institut Polytechnique de Paris France
Email: andrea.araldo@telecom-sudparis.eu

† Dept. of Electrical, Electronics and Information Engineering,
University of Catania
Italy
Email: alessandro.distefano@phd.unict.it, ad@diit.unict.it

*Abstract*—Under the paradigm of Edge Computing (EC), a Network Operator (NO) deploys computational resources at the network edge and let third-party Service Providers (SPs) run on top of them, as tenants. Besides the clear advantages for SPs and final users thanks to the vicinity of computation nodes, a NO aims to allocate edge resources in order to increase its own utility, including bandwidth saving, operational cost reduction, QoE for its users, etc. However, while the number of third-party services competing for edge resources is expected to dramatically grow, the resources deployed cannot increase accordingly, due to physical limitations. Therefore, smart strategies are needed to fully exploit the potential of EC, despite its constrains.

To this aim, we propose to leverage *service adaptability*, a dimension that has mainly been neglected so far: each service can adapt to the amount of resources that the NO has allocated to it, balancing the fraction of service computation performed at the edge and relying on remote servers, e.g., in the Cloud, for the rest. We propose *EdgeMORE*, a resource allocation strategy in which SPs express their capabilities to adapt to different resource constraints, by declaring the different configurations under which they are able to run, specifying the resources needed and the utility provided to the NO. The NO then chooses the most convenient option per each SP, in order to maximize the total utility. We formalize EdgeMORE as a Integer Linear Program. We show via simulation that EdgeMORE greatly improves EC utility with respect to the standard where no multiple options for running services are allowed.

## I. INTRODUCTION

Edge computing (EC) brings computation and data storage closer to the location where it is needed, to improve response times and save bandwidth, e.g. traffic going out from the access networks. EC is complementary to Cloud, i.e., the usual assumption is that a part of service computation is peformed at the Edge and the rest on the Cloud and similarly a part of the required data seats at the Edge and the rest on the Cloud.

We consider a Network Operator (NO), owning computational resources in its Edge network, which must decide how distribute these resources to different Service Providers (SP). The goal of the NO is to maximize its own utility, which can represent bandwidth or operational cost saving or improved experience for his users [1], [2].

A novelty of our approach is that we exploit the opportunity of running services at the Edge in different ways, i.e., the SP can balance between using more memory or more CPU, depending on the available resources, transparently to final users. For instance, in scenarios as video streaming, the SP has to deliver different representations of the same video and can choose either to store all of them, which requires a high amount of storage, or exploiting Just In Time Packaging (JITP) store just few representations and package the missing ones on-the-fly, only when needed, which saves storage but incurs more CPU usage (pag 6 of [3] and [2]). These applications show the emergence of what we call *service elasticity*: in the Edge, since resources are limited, they cannot scale with services' requirements, instead services must adapt to the available resources and run on the Cloud all the computation that cannot take place at the Edge. Partitioning limited Edge resources among third party "elastic" services is the novel core of this work. We show that by exploiting the different configurations at which SPs can run, the NO can increase its utility with respect to the classical case of one monolithic configuration per SP, as if more resources were vitrually available at the Edge than the real ones, whence the name *EdgeMORE* of our strategy.

Furthermore, we consider the distributed nature of Edge resources, which can be scattered across different nodes and the fact that services follow a microservice architectural style (Sec. V.B of [4]): a service is composed of different microservices running on *containers*. This allows fine-grained and responsive service adaptivity and resource exploitation , which makes containers attractive for Edge computing [5]. We also consider resources are multi-dimensional (memory, CPU, bandwidth, ...).

The contribution of this work is: (i) we introduce the multi-tenant multi-dimensional multiple-nodes resource allocation problem, which represents the decisions of the NO to allocates multi-dimensional resources, distributed on several nodes, among third party SPs, whose services are composed of different containers (§ III); (ii) we propose an architecture for this setting (§ IV) (iii) we provide an ILP formulation (§ V); (iv) we finally evaluate its performance in simulation (§ VI).

## II. RELATED WORK

Under the taxonomy of [5], [6], the scenario in which we study the resource allocation problem is Metro Edge Cloud and Mobile Edge Computing. Since literature on EC is vast, here we just focus on work concerning resource allocation.

*a) Multi-Tenant Resource Allocation:* Resource allocation among third party tenants is currently done in Cloud computing via pricing. However, in the Cloud resources are assumed to be infinite, so they can be granted as long as the tenant is willing to pay. At the Edge, instead, resources are limited and the NO, which owns them, want to allocate them in order to increase its own utility. Allocation of finite resources among different service providers (tenants), which compete for their consumption, has not vastly been explored in the context of EC. Some examples of this kind of problems can be found in [1] and [7], where resource is mono-dimensional (storage) and the utility is the bandwidth reduction, QoS and fairness.

*b) Resource Provisioning:* There is agreement that Edge and Cloud computing form a unique pool of resources, organized hierarchically and services can use both simultaneously. In this context, there is vast literature in resource provisioning, which regards the decision of how much resource should be deployed at the Edge nodes or at the Cloud [8], [9]. We emphasize that our resource allocation problem is different, as we assume the amount of resources at the nodes has already been decided, nodes are already deployed, and we optimize their usage by third party SPs.

*c) Network Slicing:* Network slicing consists in creating virtual network slices on top of a physical network infrastructure, whose owner has to allocate resources among slices. In this context, resources are mainly mono-dimensional (bandwidth [10]), with some exception ( [11], [12] consider also CPU).

*d) Service Adaptability:* Some work assumes, as we do, that services can run under different configuration, thus adapting to the resources provided. Services span from Federated Machine Learning [13] to video streaming [2]. Other work [14], [15] assumes that multiple configurations result in different multi-dimensional resource usage and different QoE. However, most of this work, considers one only tenant.

*e) Resource allocation for container-based EC:* The micro-service architecture is particularly suitable for resource allocation, as services can adapt to the resources available by launching/destroying the containers hosting micro-services [16]–[18].

*f) Task-oriented models:* It is important to emphasize that most of the aforementioned literature and other work [15], [19], [20] model workload as a sequence of jobs or tasks, and deals with allocating resources among them. While this models are suitable for grid computing environments, we adopt instead a service-oriented viewpoint, which we believe is more indicative of the current interactions between users and services in the Internet. Indeed, in order to live at the Edge, services consume persistent resources, which are not tightly coupled to the single user request. For instance, a content provider consumes memory to store its most popular objects, independent from the single user requests. This would not be reflected by task-oriented models.

## III. SYSTEM MODEL

We consider an edge cluster, composed of $m = 1, \ldots, M$ nodes, owned by the Network Operator (NO). These cluster nodes may be servers installed on a Central Office or machines installed in a base station. Nodes have resources of different types, e.g., RAM and CPU. We denote the types of resources as $l = 1, \ldots, L$, where $l = 1$ may indicate CPU, $l = 2$ may indicate RAM, etc. Each node has a limited amount of each resource type. We denote with $c_{l,m}$ the capacity of node $m$ in terms of resource type $l$. Node resources are virtualized so that third party Service Providers (SPs) can concurrently run their services there. Virtualization of resources is based on a container platform, like Docker, which allows to run many virtual environments, called *containers*, each hosting a third party piece of software.

SPs are denoted with $i = 1, \ldots, N$. For simplicity, and with no loss of generality, we assume that one SP runs one and only one service.[1] Each service is decomposed in a set of micro-services, each hosted in a container. Moreover, each service $i$ can run in multiple possible configuration options $j = 1, \ldots, J^i$. Each configuration option $j$ of service $i$ requires concurrently running a set of containers $z = 1, \ldots, Z^{i,j}$. Each SP $i$ declares the possible configurations under which it is capable to run and the NO decides (i) which configuration option to accept and (ii) for all the containers belonging to that option, which node they should run to. These decisions are based on utility and resource consumption.

As in [14], we assume that each configuration option $j$ of SP $i$ brings a certain utility $u^{i,j}$ to the NO. For instance, suppose SP $i$ is a video streaming service. If the NO selects a configuration option $j$ that includes containers with large memory limits, the SP will be able to cache more content and thus to serve more user requests locally, only generating traffic from/to some remote server or Cloud only for the content that is not cached at the Edge. In this case, the utility $u^{i,j}$ for the NO would be big, in terms of traffic saved.

Ideally, the NO would like to choose for each SP the option that provides the largest utility. Unfortunately, this is in general not possible, due to the scarcity of resources available in the Edge nodes. Indeed, each container consumes resources. We denote with $w^{i,j}_{l,z}$ the amount of resource of type $l$ consumed by the $z$-th container of the $j$-th option of SP $i$. Each Edge node can host different containers, from different SPs. Obviously, the sum of resource type $l$ consumed by the containers running in a certain node $m$ cannot exceed its capacity $c_{l,m}$, for any resource type $l = 1, \ldots, L$. Therefore, the NO must optimally choose one option per SP, trading off utility and resource consumption and, at the same time, optimally place the containers of the chosen options in the available Edge nodes, without exceeding their capacities.

---

[1]The terms "service" and "service provider" will thus be used interchangeably.

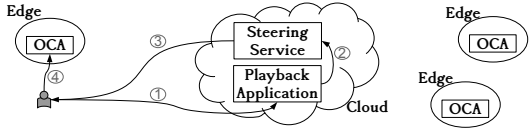

Fig. 1. Netflix architecture for OCAs CDN

Note that we do not define a per-container utility, but only the utility $u^{i,j}$ of an entire option. The rationale is that running containers individually is not useful at all. For instance, to provide an on-line gaming service, we might need a container for authenticating users, another for retrieving video frames and another to transcode them. They might all be needed together. Running the authentication server alone, may be senseless. Either the containers of an option run all or no one. Therefore, utility comes from the concurrent run of all the containers of a configuration option, and not from any single container. Since the different sets of containers that can collectively provide a service depend on the service itself, we let the SP declare the possible configuration options. Note also that we adopt a snapshot model, i.e., we assume time is divided in slots of few minutes and we focus on a single slot. At the beginning of each time slot the SPs declare their configuration options, the NO decides to accept one option per SP and assigns each container belonging to that option to one Edge node. We assume all the quantities mentioned in this section are known at the beginning of the time slot.

## IV. REFERENCE ARCHITECTURE

We now describe the architecture in which our system model can materialize. In order to ground our architecture into an existing and practical technology, we briefly outline a successful solution widely adopted by Netflix. Then, we describe our proposed architecture.

### A. An existsing implementation of Edge Computing

Netflix is one of the largest content providers. It deploys its own hardware appliances, called Open Connect Appliances (OCAs) [21], into Internet access networks. OCAs store a part of the content catalog and can serve directly a fraction of local users' requests, without generating inter-domain traffic. For this reason, NOs often accept to install this hardware in their premises. Requests are processed as in Fig. 1: (1) A user requests a video. (2) A micro-service in the Cloud selects the files to be sent to the user. (3) The steering service determines the OCAs closest to the user based on its IP address and generates a list of URLs pointing to the OCAs. (4) The user client uses the URLs list to play the video.

While this solution is currently successful and effective, it is not future-proof: it is infeasible that hundreds of SPs, like

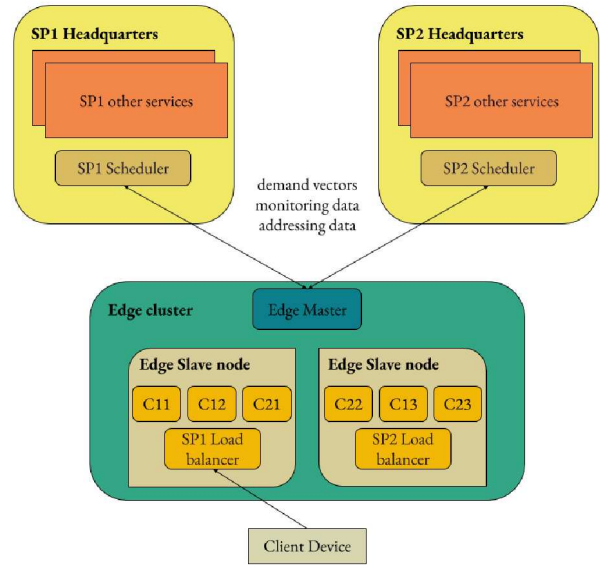

Fig. 2. Overview of the proposed architecture. SPs run part of their service in their premises or in remote Clouds, which we denote as Headquarters.

Youtube, Netflix, gaming providers, IoT providers, etc., will each install physical boxes into thousands of access networks: installing and maintaining such physical infrastructure would have an enormous cost for both SPs and NOs. Moreover, there is no physical space to host many physical boxes in the network locations at the Edge. However, the case of the OCA shows that both SPs and NOs have interest in EC, to run services at the Edge. To make EC feasible, we propose to let the NO owns the computational resources and to vitrualize them, in order to allocate them to third party SPs, acting as tenants. Each SP can then behave individually similarly to Fig. 1.

### B. EdgeMORE architecture

The components of the proposed architecture are (Fig. 2):
- *Edge slave nodes*: owned by the NO, they run the SPs' containers.
- *Edge Master*: a process controlled by the NO, responsible for (i) monitoring resource usage (e.g. using fine grained monitoring functions available in containerized environments like Kubernetes [22]); (ii) collecting the different deployment options from SPs; (iii) deciding the options to be deployed; (iv) informing the SPs about the authorized options and receiving back the containers descriptors (e.g. Dockerfile or Pods YAML); (v) running the containers in the Edge slaves. The optimization strategy of § V runs in the edge master.
- *SP Scheduler*: each SP has its own scheduler; First, it declares the set of possible configuration options to the Edge Master, specifying resource requirements and utility. After the Edge Master selects one of these options, the SP Scheduler forwards to the Edge Master the relative containers descriptor files to deploy its application at the Edge;
- *SP Load balancer*: each SP has its own load balancer; it intercepts user requests as in [23] and, based on the amount of requests served by the Edge it decides to forward the request to a remote Cloud or to handle it within the Edge [20].

*C. Edge Master workflow*

The Edge Master is the core component of the proposed architecture. Periodically, it performs the following operations.

1) It monitors the available resources and receives the set of options from the SPs schedulers; it is given as a list containing, for each option, the relevant amount of utility estimated and information on the resource requirements for each container;
2) It executes the placement algorithm to select the best option for each SP according to the collected information in point (1). The decision is sent to the SPs schedulers;
3) It receives the (chosen) option descriptors files (e.g. Dockerfiles, Smarm configurations files, Kubernetes YAML...) for the authorized options and runs these containers in the slaves nodes;
4) Finally, it communicates to SPs' load balancers the addressing data to reach the Edge internal containers. Based on the occupied resources the load balancers redirect the user requests to the Edge resources or to the Cloud.

## V. OPTIMAL RESOURCE ALLOCATION

The NO aims to maximize its overall utility, i.e., the sum of the utilities coming from all the selected options. In order to do so, the NO must concurrently take two decisions: (i) *Option selection*: the NO must select at most one configuration option per SP. (ii) *Container placement*: the NO must deploy each container of the selected options to one of the available nodes

The following is an Integer Linear Programming (ILP) formulation of the problem. The decision variables modeling the Option selection are $x^{i,j}$, which is 1 if the $j$-th option of the SP $i$ is chosen. Container placement is instead represented by the decision variables $y_{z,m}^{i,j}$, which is 1 if the $z$-th container of the $j$-th option of SP $i$ is placed on node $m$. The objective is

$$\max \sum_{i=1}^{N} \sum_{j=1}^{N_i} u^{i,j} \cdot x^{i,j} \quad (1)$$

The following constraints must be satisfied.

$$\sum_{m=1}^{M} y_{z,m}^{i,j} = x^{i,j} \quad \begin{array}{l} i = 1 \ldots N \\ j = 1 \ldots J^i \\ z = 1 \ldots Z^{i,j} \end{array} \quad (2)$$

$$\sum_{i=1}^{N} \sum_{j=1}^{J^i} \sum_{z=1}^{Z^{i,j}} y_{z,m}^{i,j} \cdot w_{l,z}^{i,j} \leq c_{l,m} \quad \begin{array}{l} l = 1 \ldots 2 \\ m = 1 \ldots M \end{array} \quad (3)$$

$$\sum_{j=1}^{J^i} x^{i,j} \leq 1 \quad i = 1 \ldots N \quad (4)$$

Equations (2) guarantee that each container $z$ of the chosen option $j$ by the SP $i$ ($x^{i,j} = 1$) is deployed ($\exists m \in \{1 \ldots M\} : y_{z,m}^{i,j} = 1$); constraints (3) guarantee that the sum of the requirements for the set of containers deployed on a node $m$ for each resource $l$ is less than the total amount of available resources in node $m$ so that these containers can actually run on the node; finally the constraints (4) guarantee that a service provider can deploy at most one option in the Edge cluster. If we have one only option per SP and a unique dimension, e.g. memory, the problem is similar to a Set-union Knapsack problem [24] and it has been solved via Dynamic Programming or via bio-inspired algorithms like bee-colony optimization [25]. If we have a single node, we can just consider for each option the total memory and the total CPU needed by all the containers composing the option. We can thus forget about the different containers and in this case we have a Multiple-Choice Multi-Dimensional Knapsack Problem (MCMDKP) [26], like in [14], although the authors do not clearly state it. Considering just one option per SP and one node, the problem reduces to a multi-dimensional knapsack problem ($l$-KP), which is a challenging problem. Methods based on the Lagrangian dual exist but difficult to apply in practice (Sec.9.2 of [27]). Moreover, Fully Polynomial Time Approximation Schemes cannot exist unless P=NP (Sec.9.4.1 of [27]), which motivates the several greedy-type heuristics proposed in the literature (Sec.9.5 of [27]). However, they cannot be directly used in our problem, which is more complicated than $l$-KP, since we need to cope with multiple options, nodes and containers. In our future work, we will explore the design of efficient heuristics to solve the problem.

## VI. NUMERICAL RESULTS

Here we present results that show how enabling service elasticity by allowing multiple configuration options to SPs notably improves the utility of the Edge. We compare the performance of EdgeMORE, computed with the ILP (§ V), with a *naive* allocation, which consists in randomly option selection and container placement. The code of the ILP in glpk and the python code to orchestrate the simulation are available as open-source [28], together with the scripts to reproduce the results presented here. The simulations run in a Intel Xeon CPU E5-4610 @ 2.30 GHz with 256GB RAM, the results are averaged across 20 runs and 95% percentiles are reported.

*a) Scenarios:* In our simulations the edge cluster consists of $M$ machines with 16 cores and $32GB$ RAM. For each simulation $N = 50$ SPs compete to gain resources in the Edge, each declaring the same number $J$ of configuration options. Each option consists of $Z = 8$ containers. The CPU and RAM required by a container are drawn from uniform random distributions with mean $\bar{w}_l$. They are expressed as dimensionless values representing CPU time for CPUs while the memory is expressed in $GB$. To obtain $\bar{w}_l$, we first fix a *load factor* $K = 1.8$ and then compute

$$\bar{w}_l \cdot Z \cdot N = K \cdot c_{l,\text{tot}}; l = \{\text{CPU,RAM}\} \quad (5)$$

where $c_{l,tot} = \sum_{m=1}^{M} c_{l,m}$ is the total amount of resource of type $l$ available at the edge. In other words, on average we allow services to request $K$ times the available resources.

The utility is also a random variable. As common in the literature [1], [7] we assume there is a concave relation between the resources used and the utility: the more resources are used by an option, the larger one should expect the utility to be, but the additional utility tends to decrease with the resources. The utility is the following function of the required resources:

$$u^{i,j} = \alpha^{i,j} \cdot \left(\frac{w_{\text{CPU}}^{i,j}}{c_{\text{CPU,tot}}}\right)^{\frac{1}{\beta_{\text{CPU}}^{i,j}}} + (1-\alpha^{i,j}) \cdot \left(\frac{w_{\text{RAM}}^{i,j}}{c_{\text{RAM,tot}}}\right)^{\frac{1}{\beta_{\text{RAM}}^{i,j}}} \quad (6)$$

where $\alpha^{i,j}, \beta_{\text{CPU}}^{i,j}, \beta_{\text{RAM}}^{i,j}$ are randomly thrown, for each option, from the random uniform distributions between 0 and 1 for $\alpha^{i,j}$ and between 1 and 5 for $\beta_{\text{CPU}}^{i,j}$ and $\beta_{\text{RAM}}^{i,j}$. Note that the formula above would be a concave increasing function if the parameters $\alpha^{i,j}, \beta_{\text{CPU}}^{i,j}, \beta_{\text{RAM}}^{i,j}$ were fixed. Choosing the parameters from a random distribution complicates the scenario. We did this on purpose since: (i) although the relation utility vs. resources can be reasonably assumed to broadly show a concave and increasing behavior, in realistic scenarios this relation may not be as "clean" as assuming a perfectly increasing and concave function; (ii) we want to check the performance of our solution in pessimistic and 'unclean' situations. For this reason, (6) is aimed to "loosely" show monotonicity and concavity. We underline that this characterization would be much more accurate if real datasets were available, which is unfortunately not the case. On the other hand, research on allocation strategies must not be paralyzed by the absence of datasets, and fortunately is not. Researchers have coped with it by proposing reasonable assumptions on the relation between resources and utility [1], [7], [14], [29] and following them, which is enough to evaluate the benefits of allocation strategies. We follow here this line.

The utility reported in the following plots is a percentage of the maximum utility, which is $N$ because $u^{i,j} \in [0,1]$ as a consequence of (6).

*b) Benefits of multiple options:* In Fig. 3 we report the effect of varying the number of options provided by each SP, assuming always $M = 8$ Edge nodes available. The utility increases with the number of options declared by SPs. Note that the classic assumption correspond to the first point of the plot, SP=1. While varying the number of options from 1 to 8 the utility has a gain almost equal to 1.6. This means that all the approaches adopted in the literature lose the opportunity to gain 60% of utility (at least in our scenario), which instead EdgeMORE can grasp by allowing SPs declare multiple options. This is the core result of the paper and also justify the name EdgeMORE: by letting SPs express their service-elasticity, it is like increasing virtually the available resources, as the ones that are available can be used better. The second plot of Fig. 3, reports that Naive uses $\sim 3.3$ times the resources of EdgeMORE, despite its poor utility, which shows that careful option selection and container placement is of paramount importance.

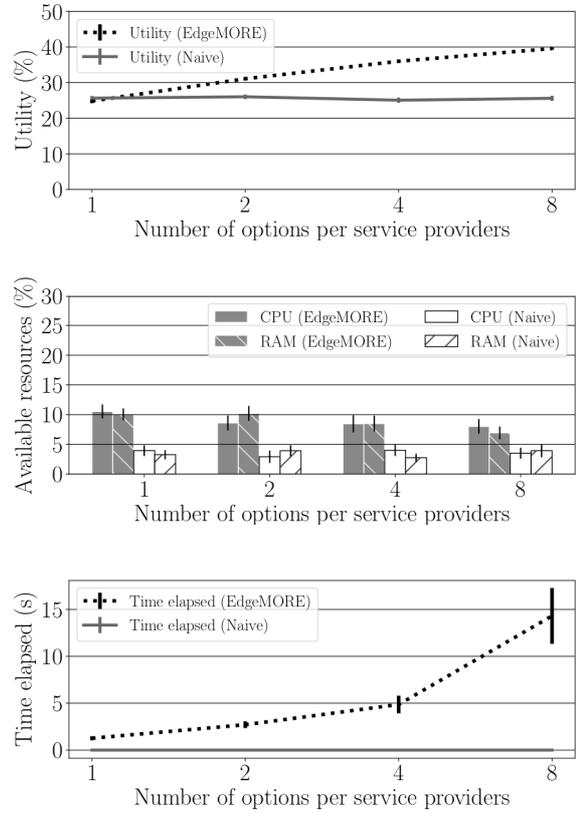

Fig. 3. Benefits of multiple options.

*c) Insensitivity to cluster scaling:* In Fig. 4 we increase the number of nodes, considering $J = 5$ options per SP. We keep $K = 1.8$, thus increasing the resources requirements proportionally (5). We also keep all the other parameters at their default values. In other words, we are testing here how the performance is affected when varying the scale of the problem, in terms of size of resources available and required. Fig. 4 shows that the utility of EdgeMORE remains unchanged with the scale of the problem. This means that the results presented here are likely to consistently appear both on tiny instances of EC as well as in larger clusters of servers available at the Edge.

*d) Computation time:* The bottom plots show that the computation time of EdgeMORE may be too high if responsive dynamic re-allocation is envisaged. This motivates to explore faster heuristics in our future work.

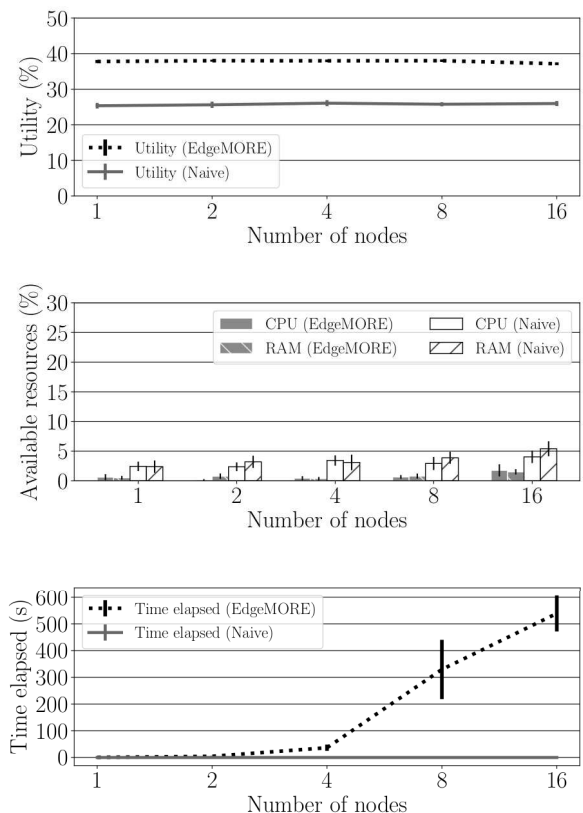

Fig. 4. Insensitivity to cluster scaling

## VII. Conclusion and future work

This paper presented EdgeMORE, a strategy for resource allocation for Edge Computing (EC), where tenants are third party Service Providers (SPs). The novelty of this work is that it exploits service elasticity: by allowing SPs to declare the different configurations (aka options) in which they can run, we show that the Network Operator (NO) owning EC resources can greatly increase utility. Relying on service elasticity is crucial in resource-constrained environments as EC. A future work will be devoted to a heuristic for the ILP and scenarios where jobs arrive in different times, exploiting a time-batched implementation of EdgeMORE. Moreover, the architecture and the strategy itself can be expanded in order to take into account different NOs (Edge roaming) and to account for inter-container communication, leveraging our previous work [30].